# A fiber integrated N-*V* diamond magnetometer compatible with commercial endoscopic systems

Satbir Singh[1], Hyunjong Lee[1], Nhu Anh Nguyen[1], Seonghyeon Kang[1], Jeong Hyun Shim[2], Sangwon Oh[3,#] , Kwang-Geol Lee[1,*]

[1]*Department of Physics, Hanyang University, Seoul 04763, Republic of Korea*
[2]*Korea Research Institute of Standards and Science, Daejeon 34113, Republic of Korea*
[3]*Department of Physics and Department of Energy Systems Research, Ajou University, Suwon, Gyeonggi, 16499, Republic of Korea*
[#]*sangwonoh@ajou.ac.kr*
[*]*kglee@hanyang.ac.kr*

Nitrogen-vacancy (N-*V*) center in diamond provides a robust, solid-state platform for magnetic field measurements at room temperature. To harness its potential in inspecting inaccessible regions, here we present a compact endoscopic configuration of an N-*V* diamond-based magnetometer. The endoscopic magnetometer was developed by integrating a large-core optical fiber with a bulk N-*V* diamond for laser excitation and photoluminescence (PL) collection. The diamond and fiber were specially shaped to enhance PL collection through the fiber. Additionally, a 3D-printed endoscope head was employed to facilitate alignment of the bias magnetic field along the N-*V* axis. A magnetic field sensitivity of approximately 3 nT/√Hz was achieved by using cw-magnetometry measurements. The endoscope diameter was restricted to 10 mm to match the dimensions of most commercial endoscopes. The magnetic field non-uniformity caused by the small separation between the diamond and the magnet in the endoscope head limited the overall sensitivity. It could be further improved to 0.85 nT/√Hz by using a magnet placed at a sufficient distance outside the endoscope head. Our endoscopic design is mechanically stable and provides additional opportunities for integrating other functionalities into the probe head as needed.

## I. INTRODUCTION.

The negatively charged nitrogen vacancy centers (N-$V^-$) in diamond have been extensively investigated for high sensitivity sensing of magnetic fields [1-4], temperature [5-7], electric field [8,9], as well as for gyroscope [10], and quantum information processing [11-13]. In magnetic field sensing, extensive work has been conducted to enhance detection sensitivity and spatial resolution. To achieve higher sensitivity, various measurement configurations based on pulse sequences [14-17] have been developed. A magnetic field sensitivity of 0.5 pT/√Hz was achieved with an ensemble of N-*V* centers [18]. For spatial resolution, single N-*V* based sensing techniques have led to nanoscale resolutions [19-21]. However, most of these high sensitivity or high resolution designs are often realized by using high-end instrumentations and thus suitable for bench-top operation in isolated environments [14-21]. This may limit the applicability of N-*V* centers to sensing magnetic fields from the sources in spatially confined or inaccessible environments. For such purposes, a portable magnetometer with a mobile probe head is highly desirable.

For the realization of portable N-*V* diamond-based magnetometers, fiber-integrated configurations have been developed [22-26]. In these configurations, the N-*V* diamond is generally integrated on the apex of an optical fiber, and microwaves are delivered via a nearby attached loop or a wire. Multiple approaches, such as the use of the cubo-octahedron shaped microdiamond [22], the incorporation of a parabolic condenser lens and fiber collimator [23], and the use of photonic crystal fiber [25], have been considered in previous works. The magnetic field sensitivity in the sub-nT range has been reported for such fiber-integrated magnetometer configurations [22,23]. While a fully integrated, portable configuration with sub-nT magnetic field sensitivity has also been developed [27], its use in typical endoscopic applications may be limited due to design constraints.

A compact endoscopic configuration having all the essential components fully integrated into a probe head is desirable for practical applications ranging from the clinical to the industrial sector. In the present study, a simpler and mechanically robust endoscopic magnetometer is developed for applications in difficult-to-access regions. The endoscopic configuration is based on the integration of a bulk ensemble N-*V* diamond and a large-core (diameter 1.5 mm) optical fiber. Both the diamond and the fiber were shaped to improve the PL collection through the fiber. A 3D-printed endoscopic head was employed to integrate a permanent magnet to apply a bias field and a microwave delivery coil in the endoscopic probe. The details of the endoscopic design are presented in the Methods section. The magnetic field sensitivity of around 3 nT/√Hz was obtained using continuous wave (CW) magnetometry measurements. Although the obtained sensitivity is lower than for some of the previously reported fiber-based systems, the primary objective of this study is to explore the feasibility of a compact and simpler endoscopic configuration.

## II. METHODS

### A. Construction of endoscope

The endoscopic magnetic sensor developed in this study employs a (100) cut single-crystal diamond (DNV-B14, Element Six) with an N-*V* center concentration of approximately 4.5 ppm. The N-*V* diamond was attached to a multimode optical fiber (FP1500ERT, Thorlabs) with a 1.5 mm core diameter and a numerical aperture (NA) of 0.50. This fiber serves as the main body of the endoscope for excitation of N-*V* centers and collection of photoluminescence (PL). The magnetic field sensitivity of the N-*V* diamond magnetometer fundamentally depends on the collected PL intensity [1-4]. Therefore, both the diamond and the optical fiber were specifically shaped to improve PL collection through the fiber. The refractive indices of the core and cladding of the used large-core fiber are specified to be 1.458 and 1.365, respectively. This corresponds to critical angles of 69.5° for the core/cladding interface and 43.3° for the core/air interface. After attachment of N-*V* diamond to the apex of the fiber, the emission angle of the PL signal from the diamond (n~2.42 @ 640 nm) into the fiber core ranges between 0 to 90°. Thus, the inefficient total internal reflection at certain angles in the fiber section can lead to a certain amount of PL signal loss. This problem can be remedied by shaping the fiber facet into a curved geometry with a gradually varying angle from 0° to 43.3° as shown in Fig. 1(a). A combination of diamond lapping/polishing sheets (Thorlabs) and a rotational grinding machine was used for precisely curving the bare optical fiber at the desired angle. The thin cladding layer was also removed during the process, allowing for a lower critical angle condition at the core/air interface. The fiber apex diameter was reduced from 1.5 mm to 1.0 mm. Using the same analogy, the diamond slab (1.5 × 1.5 × 0.5 mm$^3$) was also polished into a truncated cone geometry with a 1.0 mm base diameter, 0.5 mm height, and 0.5 mm top diameter. The semi-vertical angle was ~ 25° (critical angle for total internal reflection at diamond/air interface is 24.6°), as shown in Fig. 1(b). This design aims to reduce the trapping of the PL signal inside the high-refractive-index diamond and to efficiently direct the PL into the fiber. This shaping work was outsourced to Almax easyLab, Belgium. The truncated cone N-*V* diamond was affixed to the apex using Norland optical adhesive NOA 146H. A non-magnetic coaxial cable (UT-047C-TP-LL, Amphenol CIT) was used for delivering microwave excitations to the N-*V* diamond for transitions between $m_s = 0$ and $m_s = \pm 1$ spin states.

The diamond/fiber assembly was inserted into an endoscopic instrumental channel tube with external and internal diameters of ~4 mm and ~3 mm, respectively. For demonstration purposes, a commercially available endoscopic clamp tube (EG-590WR, FUJINON Gastroscope) was used as the instrumental channel. Note that conventional endoscopes facilitate an instrumental channel with a diameter of ~3 mm (outer diameter is ~ 4 mm). The fiber was secured inside the tube using heat-shrink tubing, which was tightened sufficiently to prevent unintentional movement while still allowing manual rotation and repositioning. The inner conductor of a semi-rigid coaxial cable was wrapped around the instrumental channel tube precisely at the N-*V* diamond position. The other end of the inner conductor was grounded by connecting to the outer conductor. In this configuration, the instrumental channel tube protects the microwave wire from accidental contact with the diamond, providing a stable configuration.

The alignment of a bias magnetic field along one of the four N-*V* axes is a typical way of isolating a single N-*V* orientation [2-4]. The component of the measured magnetic field along this axis can be extracted from the shifts in the optically detected magnetic resonance (ODMR) frequency. For this purpose, a custom endoscopic head was specially designed and fabricated using an in-house 3D-printing, as shown in Fig. 1(c). The endoscopic head has a channel for placing the instrumental channel tube. In the head, there is a provision for placing a permanent magnet (3 mm diameter, 1 mm width, N50, Neodymium magnet) aligned toward a pre-selected N-*V* diamond position. The diameter of the endoscopic head was deliberately restricted to 10 mm, which is the typical size of clinically accepted endoscopes, for instance, Endoeye HD II 10 mm, Olympus. This enables the design's applicability in a wide range of applications. The vacant space in the endoscope head can be further managed to incorporate additional channels for monitoring, cleaning, etc.

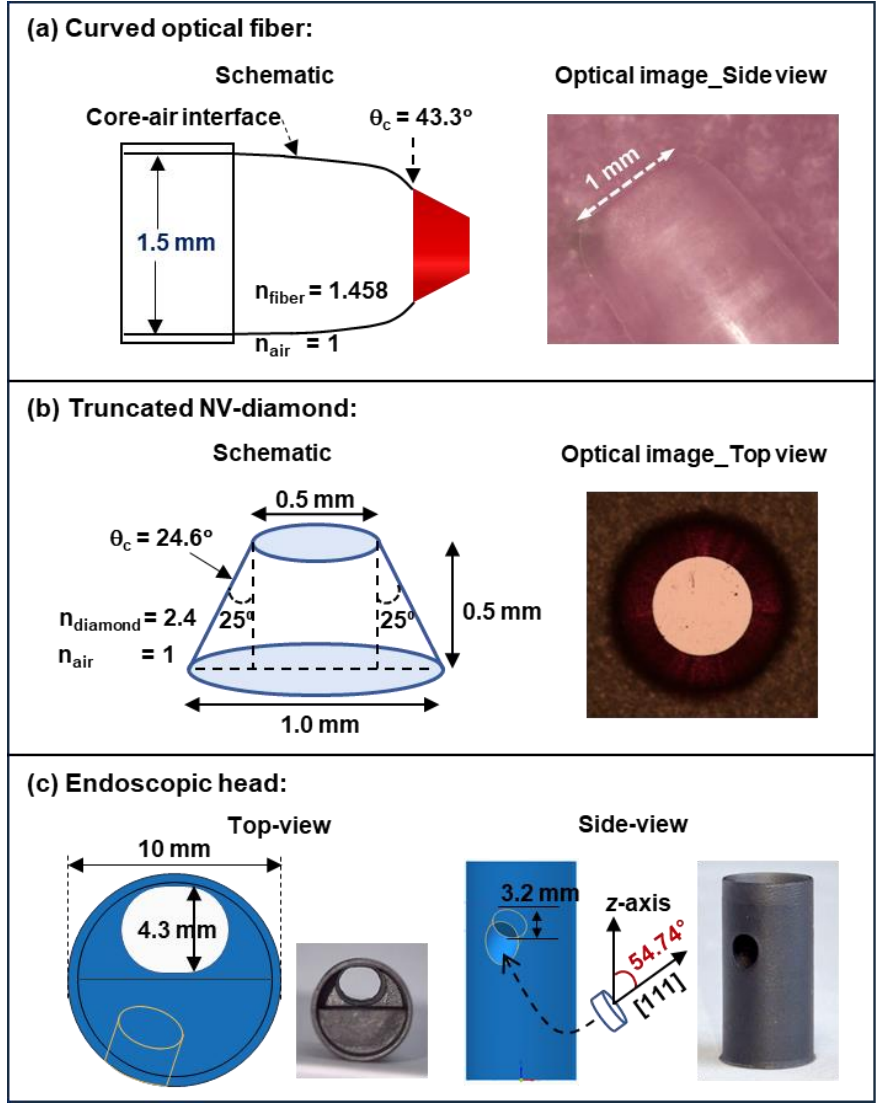

FIG. 1 (a) Schematic (left) and optical microscope image (right) of the shaped large-core fiber. The fiber core has a refractive index of 1.458, with a critical angle of 43.3 ° at the core/air interface. The fiber apex was polished in-house using a grinding rotator motor, gradually varying the angle from 0 ° to 43.3 ° near the tip to minimize PL loss. The surface was then finished with a polishing film. (b) Bulk N-*V* diamond polished into a truncated cone (Almax easyLab, Belgium). (c) Design of the custom 3D-printed endoscopic head (top and side views), showing dimensions and magnet configuration for the bias field, along with corresponding photographs.

As the endoscopic design demands a narrower diameter (~10 mm), the distance between the N-*V* diamond and a permanent magnet is constrained. Alternative approaches, such as attaching a combination of small magnets directly to the fiber near the diamond, were also explored to isolate the N-*V* axis. However, in our trials, these methods proved to be technically challenging and less effective for achieving precise control of the bias magnetic field alignment. However, this approach has recently been investigated in another study, showing the growing interest in the development of N-*V* diamond-based endoscopic magnetometer [28]. In contrast, our 3D-printed endoscopic head allows a ring-shaped permanent magnet to be positioned in a dedicated slot angled at 54.74 ° to the vertical axis, corresponding to the N-*V* orientation in the diamond lattice. The fiber was fitted comfortably within the endoscope, enabling precise rotation and axial translation for fine alignment of the magnetic field along the N-*V* axis. The diamond-fiber assembly was also carefully positioned and rotated inside the instrumental channel tube to align the N-*V* axis with the magnetic field. This controllable rotation and translation of the fiber-integrated diamond provides flexibility in selecting the N-*V* axis, in addition to the mechanical stability provided by the heat-shrink tubing around the fiber.

The back end of the optical fiber was used to deliver the 532 nm continuous-wave (cw) excitation laser to the N-*V* diamond and to collect the PL signal, unless otherwise mentioned. All other optical components, except the excitation laser, were integrated into a portable setup (30 cm $\times$ 30 cm), as depicted in Fig. 2. The excitation laser was coupled into the setup through a fiber-coupling arrangement. The 532 nm excitation and the PL signal were separated using a dichroic mirror, and a 650 nm long-pass filter was used for selecting the PL of N-*V*⁻ centers. The microwave was delivered by connecting the amplified output to the semi-rigid coaxial cable through an SMA connector (not shown). This configuration allows the optical setup to be movable to remote locations.

### B. Instrumentation

A continuous-wave 532 nm laser (CNI-MGL-III-532 nm) was used to excite the N-*V* diamond. The laser power at the portable optical setup was 300 mW. Two identical photodetectors (PDA100A, Thorlabs) operated at the same gain settings were used to detect the PL and reference laser signals for balanced measurement. Microwaves were produced with a Hewlett-Packard E4421B signal generator and amplified by a Mini-Circuits ZHL-16W-43-S+ power amplifier. To protect the microwave source, the line included a microwave circulator/isolator and 50 Ω terminations. The internal low-frequency function generator built into the microwave signal generator provided the low-frequency signal for frequency modulation. A digital lock-in amplifier (SR850, Stanford Research Systems) was used for lock-in detection of the frequency-modulated signal. The balanced configuration of the lock-in amplifier was used to suppress laser-source-induced noise. The time constant of lock-in was set to 1 ms, corresponding to a 125 Hz bandwidth at 12 dB/octave filter settings. The signal was maximized using the in-phase output (X) of the lock-in, and digitized using an NI BNC-6363 data acquisition board. A homemade LabVIEW (National Instruments) program was used for synchronization and measurement control.

### III. RESULTS AND DISCUSSION

The optical response of the shaped diamond–fiber assembly was evaluated using a home-built confocal microscope. For this, as shown in Fig. S1 (Supplemental Material [29]), the N-*V* diamond attached to the fiber was directly excited with an air objective lens (0.6 NA), and the PL signal collected at the fiber end was ~12 times stronger than that detected through the same objective. This demonstrates efficient PL coupling from the truncated-shaped diamond into the curved fiber. Ray-tracing simulations (see Fig. S2 in the Supplemental Material [29]) further confirmed this effect, indicating ~23% PL collection efficiency, compared with ~2% for a diamond slab with a 0.6 NA air objective and ≤10% with a 1.45 NA oil-immersion objective [30]. The enhanced collection achieved with the shaped diamond–fiber assembly highlights the advantage of our design. Moreover, the optical fiber provides a flexible and portable medium for both excitation and collection, paving the way for compact endoscopic N-*V* diamond magnetometers.

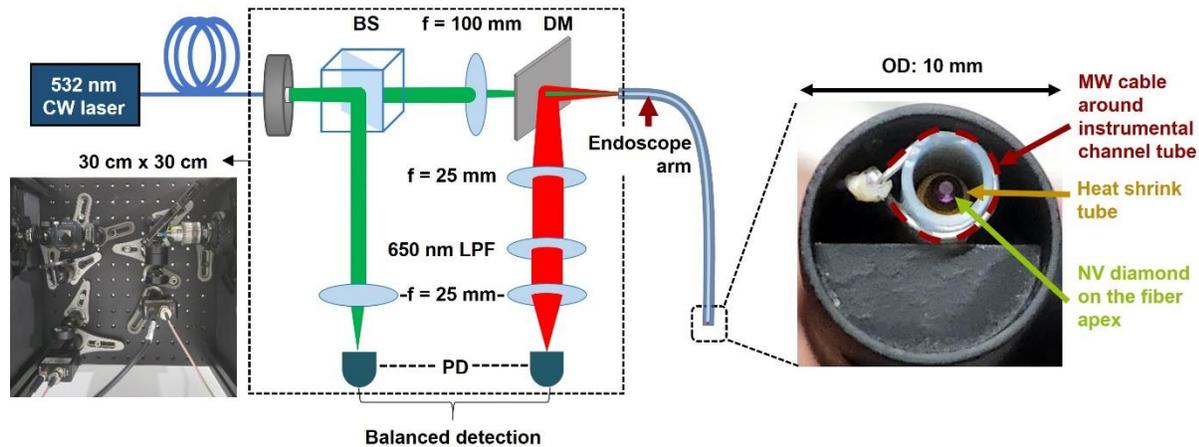

FIG. 2. System configuration of the developed N-*V* diamond-based endoscopic magnetometer, showing the optical measurement setup and the endoscopic head. BS: beam splitter, DM: dichroic mirror, LPF: long pass filter, PD: photodiode, OD: outer diameter, MW: microwave.

To test the performance of the endoscope-designed magnetometer, cw-ODMR spectrum for the magnetic field aligned along the N-*V* axis was measured by using the lock-in detection method. The microwaves were frequency modulated for the lock-in detection using a sine wave low-frequency reference signal. The frequency modulation depth was optimized to maximize the slope in the lock-in signal corresponding to the [111] axis. A representative ODMR spectrum, measured at a modulation frequency of 5 kHz and a modulation depth of 8.5 MHz, is shown in Fig. 3(a). The spectrum shows a clear ODMR signal for a well-isolated [111] axis. The linear region with maximum slope for $m_s = 0$ to $m_s = -1$ resonance is marked in red.

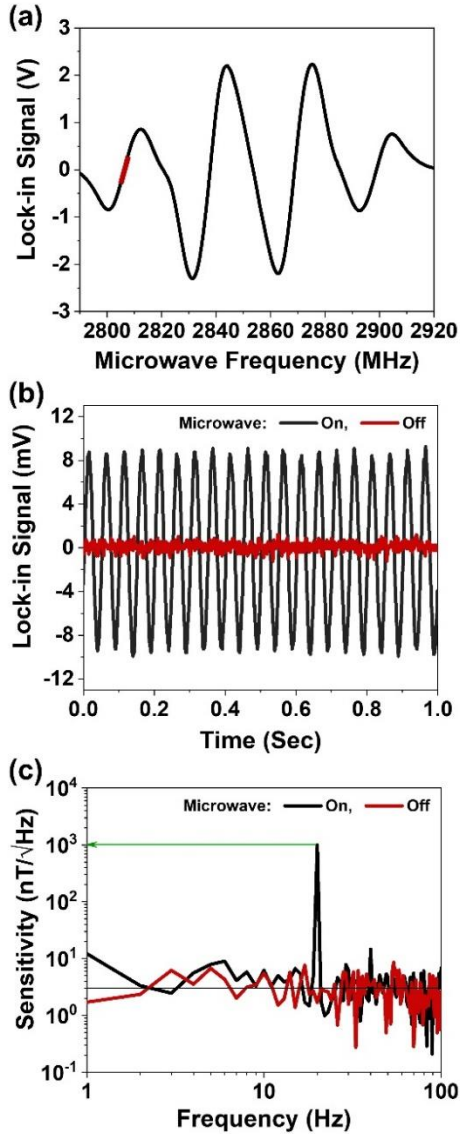

FIG. 3. (a) ODMR spectrum obtained with a bias magnetic field aligned along the N-*V* axis using a permanent magnet placed in the endoscopic head. (b) Time trace recorded in the linear region of the lock-in spectrum (red in (a)) under an applied test magnetic field with a projection strength of 1 µT RMS at 20 Hz along the N-*V* axis. (c) Magnetic field sensitivity derived from FFT analysis of the time trace in (b), showing an average noise floor of ~3 nT/√Hz. Black and red curves in (b) and (c) correspond to the cases with microwave turned on and off.

The magnetic field sensitivity of the developed endoscope was evaluated by applying a known AC test magnetic field. A home-built Helmholtz coil was used to generate a magnetic field of 1.73 µT RMS at a frequency of 20 Hz along the [100] crystallographic direction at the N-*V* diamond position. The component of this field projected onto the [111] axis was 1 µT RMS. Prior to these measurements, the Helmholtz coil was calibrated by comparing the theoretically calculated magnetic field, based on the Helmholtz coil configuration with the magnetic field values inferred from shifts in the ODMR resonance frequencies. As shown in Fig. 3(b), a time trace was recorded at a sampling rate of 1kS/s over 1 sec, within the linear region of the lock-in signal highlighted in red in Fig. 3(a). The time domain signal was analyzed using a Fast Fourier Transformation (FFT) to obtain the RMS amplitude spectral density. A distinct spectral peak at 20 Hz, corresponding to the applied test field of 1 µT RMS, was observed, as shown in Fig. 3(c). This analysis indicates an average magnetic field noise floor of approximately 3 nT/√Hz. For comparison, the magnetic field sensitivity with a diamond slab with the same specification (DNV-B14) was measured using a home-built confocal configuration, yielding a noise floor ~1 nT/√Hz. The lower sensitivity in the endoscope case, compared to the confocal measurements, is attributed to broadening of the ODMR line-width due to the magnetic field non-uniformity over the illuminated area of the diamond. This suggests that sensitivity could be improved by optimizing the configuration of the permanent magnet within the endoscope head, an approach that will be explored in future work.

The effect of magnetic field non-uniformity in the diameter-constrained endoscopic magnetometer was further evaluated. For this, a bias magnetic field along the N-*V* axis was applied using a larger diameter permanent magnet placed at a sufficiently large distance, instead of the closely placed endoscopic magnet. This configuration was assumed to provide a uniform magnetic field at the N-*V* diamond. In the following discussion, this configuration is referred to as configuration (B), while the original design with the magnet integrated into the endoscope head is referred to as configuration (A). The ODMR spectrum measured with configuration (B) using the lock-in detection method is shown in Fig. 4(a). The hyperfine splitting was clearly observed, which was absent in the ODMR spectrum for configuration (A). The optimized modulation depth was ~ 0.6MHz. This suggests that the non-uniform magnetic field at the N-*V* diamond in the configuration (A) of the endoscope causes linewidth broadening, thereby limiting the sensitivity.

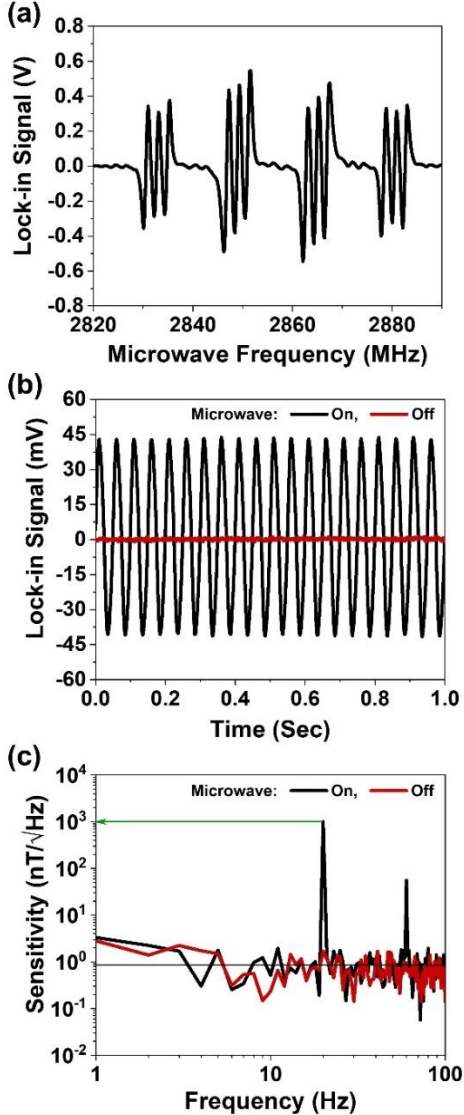

FIG. 4. (a) ODMR spectrum with a bias magnetic field aligned along the N-*V* axis using a magnet placed at a sufficient distance to provide a uniform field at the N-*V* diamond. (b) Time trace recorded in the linear region of the lock-in spectrum under the same test magnetic field as in Fig. 3. (c) Magnetic-field sensitivity derived from FFT analysis of the time trace in (b), showing a noise floor of ~0.85 nT/√Hz. Black and red curves in (b) and (c) correspond to the cases with microwave turned on and off.

Further, the same test field at 20 Hz frequency with 1 μT RMS projected along the [111] axis was applied. As shown in Fig. 4(b), a time trace was recorded at a sampling rate of 1kS/s over 1 sec, within the linear region of the lock-in signal for $m_s = 0$ to $m_s = -1$ transition. From the FFT analysis of the time trace in Fig. 4(b), an average magnetic field noise floor ~0.85 nT/√Hz was obtained, as shown in Fig. 4(c).

These results show a significant improvement in the sensitivity compared to configuration (A). It is worth noting that for configuration (B), further improvement in the sensitivity (~ 4 times) can be expected through simultaneous driving of hyperfine transitions and double resonance excitation [31]. Configuration (B) can be suitable for test environments that allow the use of externally applied bias field. However, configuration (A) remains advantageous in scenarios where the use of an external magnet is restricted, as is typically the case in endoscopic applications.

## V. CONCLUSIONS

An endoscopic configuration of an N-*V* diamond-based magnetometer was developed. A bulk diamond with an N-*V* concentration of ~4.5 ppm was attached to the apex of a large-core optical fiber, which served for both excitation and PL collection. To maximize PL collection, the diamond was polished into a truncated-cone geometry with a semi-vertical angle corresponding to the critical angle (24.6°) of the diamond/air interface, and the fiber facet was shaped into a curved geometry based on the core/air critical angle (43.3°). The diamond–fiber assembly was inserted into a commercially available endoscopic clamp tube (4 mm outer diameter), occupying the instrument channel of conventional endoscopes. A custom 3D-printed head was designed to accommodate the channel tube and to include a slot for a permanent magnet, enabling application of a bias magnetic field. Manual control of the fiber–diamond assembly allowed rotation and translation for aligning the N-*V* axis with the magnetic field. The total diameter of the endoscopic head was restricted to 10 mm, ensuring compatibility with clinically used endoscopes. A magnetic field sensitivity of ~3 nT/√Hz was achieved using cw-magnetometry. When the bias field was applied with a magnet placed externally to the endoscope, the sensitivity improved to ~0.85 nT/√Hz. Both configurations demonstrated here provide practical utility for magnetic-field measurements in otherwise inaccessible regions.

## ACKNOWLEDGMENTS

This work was supported by the Institute of Information and Communications Technology Planning & Evaluation (IITP) grant funded by the Korean government (MSIT) (Nos. RS-2022-II221026, RS-2025-02215576), and Global-

Learning & Academic research institution for Master's PhD students and Postdocs (G-LAMP) Program of the National Research Foundation of Korea (NRF) grant funded by the Ministry of Education (No.RS-2023-00285390).

# Supplementary Material

# A fiber integrated N-*V* diamond magnetometer compatible with commercial endoscopic systems

**Satbir Singh[1], Hyunjong Lee[1], Nhu Anh Nguyen[1], Seonghyeon Kang[1], Jeong Hyun Shim[2], Sangwon Oh[3,#], Kwang-Geol Lee[1,*]**

[1]Department of Physics, Hanyang University, Seoul 04763, Republic of Korea

[2]Korea Research Institute of Standards and Science, Daejeon 34113, Republic of Korea

[3]Department of Physics and Department of Energy Systems Research, Ajou University, Suwon, Gyeonggi, 16499, Republic of Korea

#sangwonoh@ajou.ac.kr

*kglee@hanyang.ac.kr

I. PL collection efficiency of the shaped fiber-diamond assembly _ experiment

The effect of fiber and diamond shaping on the PL collection through the fiber was investigated. For comparison purposes, a conventional confocal configuration for measuring optically detected magnetic resonance (ODMR) setup was considered. The shaped diamond attached to the curved fiber was directly excited with a 532 nm laser using a 0.6 NA air objective. The PL was collected through the same objective lens (Fig. S1 (a)), and also from the opposite end of the fiber (Fig. S1 (b)). As shown in Fig. S1 (c), the PL detected at the fiber end was ~12 times more than that collected through the objective lens. This result indicates efficient PL coupling into the fiber, resulting in an enhanced signal collection through the fiber.

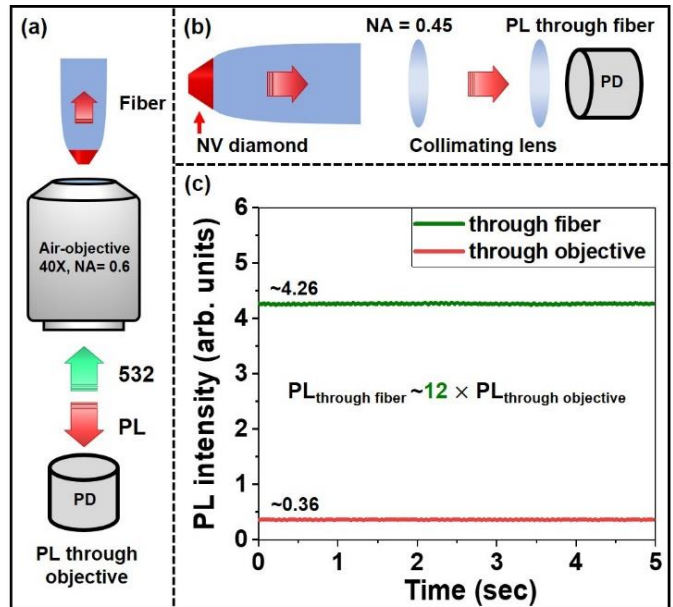

FIG. S1. Comparison of PL collection (a) through the excitation objective lens and (b) through the shaped fiber. (c) The measured PL intensity at the fiber end is approximately 12 times higher than that collected through the objective.

## II. Ray tracing simulations

The PL collection efficiency from the truncated cone shaped diamond into the curved fiber was estimated by performing ray tracing simulations with the COMSOL Multiphysics ray optics module. The diamond used in the endoscope construction was DNV B14, with an NV concentration of 4.5 ppm uniformly distributed throughout the entire volume. In this simulation, the same situation is assumed. Within the truncated-shaped diamond, randomly distributed 2000 rays with isotropic radiation pattern were considered. Approximately 23% of the total emitted power was collected at the back end of the fiber as shown in Fig. S2. This value is significantly higher than the typical PL collection efficiencies from the diamond slab in confocal configuration which is ~2% with 0.6 NA air objective, and ≤10% for 1.45 NA oil-immersion objective [1]. This result demonstrates improved collection efficiency in the shaped diamond fiber configuration.

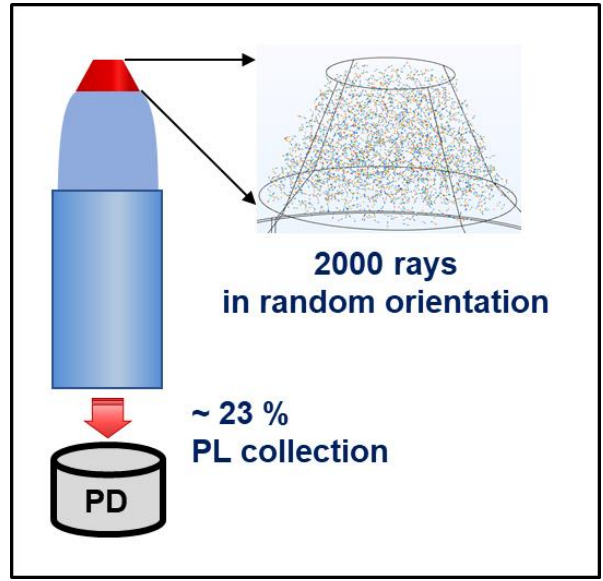

FIG. S2. Ray tracing simulations suggests ~23 % collection efficiency through the fiber